\documentclass[aps, reprint,prl,superscriptaddress,nofootinbib]{revtex4-1}
\usepackage{dcolumn}
\usepackage[utf8]{inputenc}
\usepackage{amsmath}
\usepackage{amsfonts}
\usepackage{amssymb}
\usepackage{graphicx}
\usepackage{hyperref}
\usepackage{xcolor}
\usepackage{adjustbox}

\newcommand{\CNOT}{\textsc{cnot}}
\newcommand{\CPHASE}{\textsc{cphase}}
\newcommand{\SWAP}{\textsc{swap}}

\newcommand{\R}{\textsc{r}}
\newcommand{\ket}[1]{\ensuremath{|{#1}\rangle}}
\newcommand{\bra}[1]{\ensuremath{\langle{#1}|}}

\newtheorem{teorema}{Theorem}

\usepackage{tikz}
\usetikzlibrary{quantikz}
\begin{document}

\title{Temporal witnesses of non-classicality and conservation laws}

\author{Giuseppe Di Pietra}
\email{giuseppe.dipietra@physics.ox.ac.uk}
\affiliation{Clarendon Laboratory, University of Oxford, Parks Road, Oxford OX1 3PU, United Kingdom}
\affiliation{Dipartimento di Scienza Applicata e Tecnologia, Politecnico di Torino, Corso Duca degli Abruzzi 24, I–10129 Torino, Italy}

\author{Chiara Marletto}
\affiliation{Clarendon Laboratory, University of Oxford, Parks Road, Oxford OX1 3PU, United Kingdom}
\affiliation{Centre for Quantum Technologies, National University of Singapore, 3 Science Drive 2, Singapore 117543, Singapore}

\date{\today}%

\begin{abstract}
A general entanglement-based witness of non-classicality has recently been proposed, which can be applied to testing quantum effects in gravity. This witness is based on generating entanglement between two quantum probes via a mediator. In this paper we provide a ``temporal" variant of this witness, using a single quantum probe to assess the non-classicality of the mediator. Within the formalism of quantum theory, we show that if a system $M$ is capable of inducing a coherent dynamical evolution of a quantum system $Q$, in the presence of a conservation law, then $M$ must be non-classical. This argument supports witnesses of non-classicality relying on a single quantum probe, which can be applied to a number of open issues, notably in quantum gravity or quantum biology. 
\end{abstract}

\maketitle

%\pacs{}

\section{Introduction}

Quantum theory is in principle universal: there are no known principles setting exact limits to the domain of applicability of its unitary dynamics. Hence in principle the descriptors of all physical systems are ``q-numbers" (i.e., non-commuting operators) \cite{DEWITT}; quasi-classical behaviour emerges from this quantum structure via decoherence \cite{WALLACE}; and quantum-interference effects can in principle be observed at arbitrarily large scales, just like Schrödinger and Wigner imagined in their thought-experiments \cite{SCH, WIG, DEU}. Unitary quantum theory provides therefore a completely self-consistent picture of the world, applicable to measuring devices, observers, and other macroscopic systems, such as living entities. Nonetheless, the universality of quantum theory is still questioned to this day. Numeorus arguments have been proposed against its universality: for instance, some speculate that macroscopic observers may have  ultimately to be classical \cite{BOHR}; others claim that specific systems, such as the gravitational field, are best described by a classical or semi-classical model \cite{PENROSE, GRW}.

In this context, it is of the essence to find experimental tests to rule out classical models for gravity conclusively, under minimal assumptions. A particularly promising approach to these tests has been recently proposed, with a novel ``witness of non-classicality'' based on the entangling power of gravity. In particular, an argument was proposed showing that if a system $M$ can mediate (by local means) entanglement between two quantum systems, $Q$ and $Q'$, then it must be non-classical \cite{MAVE, MRVPRD}. ``By local means" here indicates a  protocol, detailed in \cite{SOUG, MAVE, MRVPRD}, where $Q$ and $Q'$ must not interact with each other directly, but only via the mediator $M$. Also,  ``non-classicality'' is a theory-independent generalisation of what in quantum theory is expressed as ``having two distinct physical variables that do not commute". Informally, being non-classical means having two or more physical variables that cannot simultaneously be measured to an arbitrarily high degree of accuracy \cite{MRVPRD}. This argument offers a broad theoretical basis for recently proposed experiments that can test quantum effects in gravity at the laboratory scale  \cite{MAVE, SOUG}, and for any other experiment that (beyond the gravity case) intends to show that some system $M$ is non-classical. 

This witness relies on the capacity of a system $M$ to generate entanglement between two space-like separated degrees of freedom; hence a natural question to consider is whether one could generalise the witness by exploiting the well-known correspondence between spatial and temporal entanglement \cite{Vedral, Brukner, REF1, REF2, REF3, Leggett}. 
This generalisation may shed new light on the meaning of the so-called \textit{locality in time}, giving a novel understanding of the connection between space and time in nature. Moreover, it may lead to simpler experimental schemes, relying on observing the dynamical evolution of a single probe, rather than on observing entanglement of two probes.

\begin{figure}
    \centering
    \includegraphics[width=\columnwidth]{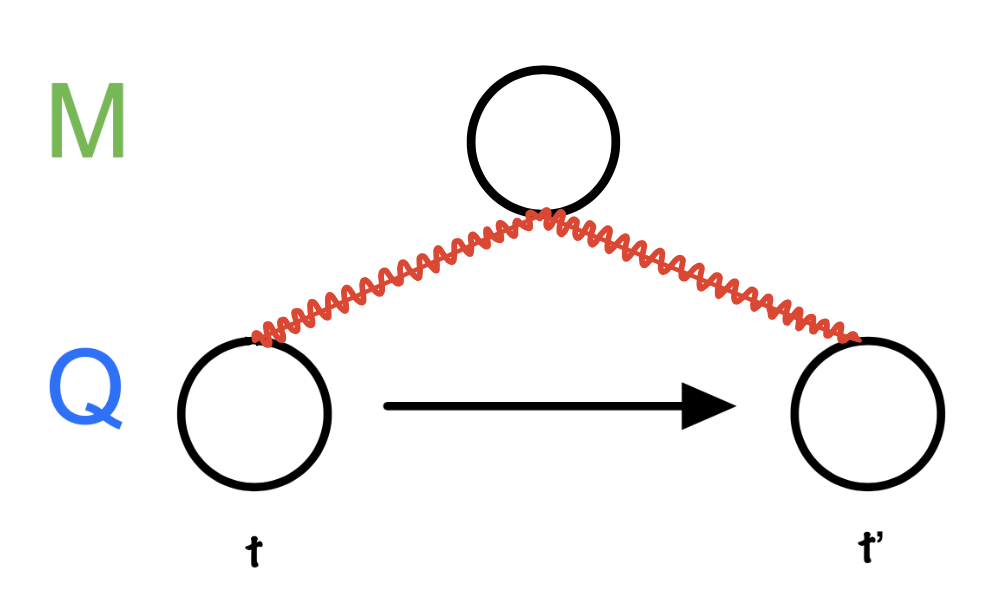}
    \caption{Schematic representation of the temporal witness of non-classicality. $Q$ is the probe and $M$ is the system whose non-classicality is to be assessed, by testing its ability to induce a non-trivial quantum-coherent evolution of $Q$.}
    \label{fig:scheme}
\end{figure}

In a manner reminiscent of Leggett-Gaarg inequalities \cite{Leggett}, we shall consider a {\sl single} quantum probe $Q$ measured at two times, and a mediator $M$ that induces a quantum-coherent time-evolution of $Q$ between these two times (see Fig.\ref{fig:scheme}). We shall then find sufficient conditions under which we can conclude that if $M$ can induce a quantum-coherent evolution of $Q$, then $M$ is non-classical. As we shall explain, a key assumption will be the conservation of a global observable of the joint system of $Q$ and $M$. We propose that this conservation law is the temporal equivalent of the locality constraint in the entanglement-based version of the argument. This idea bodes well with the well-known relation between temporal and spatial entanglement. We note that the single-mass experiment proposed in \cite{HOW} to test quantum effects in gravity, aims at discriminating between a particular quantum model of gravity and its semi-classical version. Here instead we propose a general witness of non-classicality that can rule out a vast class of classical models for the mediator $M$ obeying our general assumptions, not just a particular classical model. This result is an important step to give the temporal witness of non-classicality the same generality as the spatial entanglement-based witness. Moreover, our argument applies to {\sl any} physical system $M$ provided its interaction with the quantum probe $Q$ satisfies our assumptions, for instance a macroscopic system as a polymer. It therefore can provide theoretical support to the claim that the quantum mechanisms are relevant even at a macroscopic scale including living systems — an issue highly relevant for quantum biology.

\section{Temporal witness of non-classicality}

Following \cite{MRVPRD}, we shall define a system as ``non-classical" if it has at least two distinct physical variables that cannot be measured to arbitrarily high accuracy by the same measuring device. In the framework of quantum physics, on which our argument will rely at this stage, these two ``incompatible'' variables are what we call ``non-commuting'' observables. The definition we use for non-classicality is a special case of the information-theoretic definition used in \cite{MAVE}: a physical system is classical when it has one observable, defined as a set of perfectly copiable attributes \cite{MRVNPJ, MRVARXIV}. Using the formalism of quantum theory, this definition can be formalized as requiring that the observable of the system in question are all commuting with each other. This notion of classicality interestingly differs from others, for instance as system being classical if it is prepared in a coherent state or if it is a decoherent channel \cite{CLAS}.

We shall consider two systems: $Q$, a quantum system; and $M$, a system that we initially assume to have only one observable $Z_M$: the set of classical states is included in the span of its eigenstates, which forms a vector space. $M$ may or may not be non-classical: our theorem will indicate a way to test its non-classicality under two assumptions.
Our assumptions will be: (i) Conservation of a global observable on $M$ and $Q$, which (on $M$'s side) is a function of the `classical' variable $Z_M$ only ; (ii) The formalism of quantum theory.  We expect that it is possible to relax the latter assumption; at this stage, this argument represent a witness of the \textit{quantum} nature of the mediator $M$. Among other things, assumption (ii) implies the interoperability of information \cite{DEMA}, which we conjecture is needed to develop a more general argument, in parallel with the general argument supporting the entanglement-based version of the witness \cite{MRVPRD}. 

Without loss of generality, for simplicity we will also assume that $Q$ is a qubit, with $Z_Q$ representing its computational basis, and  $M$ is a bit, so all their observables are two-fold.  

Let us now define the {\sl witnessing task} as the task where $Q$ evolves from an eigenstate of the conserved quantity $Z_Q$ to an eigenstate of another observable $X_Q$ that does not commute with $Z_Q$. This corresponds to creating quantum coherence on the system $Q$, in the basis where $Z_Q$ is diagonal.

We will now show that if the witnessing task can be achieved by letting $Q$ interact with $M$ only, then $M$ must be non-classical. We shall prove this result first by assuming an additive conservation law, and then a non-additive conservation law. 

\subsection{Conservation of an additive quantity}
Assumption (i) is formally expressed as the requirement that the quantity $Z_M+Z_Q$ is conserved.
Specifically, we will require that any allowed dynamical transformation $U_{MQ}$ acting on the joint system $M\oplus Q$ should satisfy: 
\begin{equation}
[U_{MQ}, Z_M+Z_Q]=0 \label{CON}
\end{equation}

We can readily see that this constraint together with the assumption that $M$ is a classical bit, with $Z_M$ being its only observable, leads to the impossibility of the witnessing task:  $Q$ cannot evolve if the conserved quantity is initially sharp. Note that the conservation law is enforced by requiring that the only allowed unitaries have the property that the operators $Z_M$ and $Z_Q$ are left unchanged. Hence, we cannot exploit $Z_M$ to ``read" the initial state of the probe $Q$ and therefore to induce the coherent evolution required for the witnessing task. Moreover, working with operators when enforcing the conservation law makes our argument independent of the initial state for the mediator $M$. This is an interesting point since we need not have direct control on the initial state of the system $M$ during the protocol we envisage.

For the only allowed unitary by \eqref{CON} must be a function of a Hamiltonian whose most general form is $\alpha Z_Q+\beta Z_M+\gamma Z_QZ_M$, with $\alpha, \beta, \gamma$ real constants. Such a unitary cannot make $Z_Q$ unsharp, if it is sharp initially. 
Hence if $Z_Q$ is made unsharp by interacting with $M$ only, and \eqref{CON} is satisfied, $M$ must be non-classical. 
Note that this argument does not rely on a specific initialisation for the mediator $M$.

It is also possible to generalise this result to a quantum channel operating on $Q$ and $M$, under the following assumptions. One, that the channel can be extended to a unitary transformation on three systems, $Q$, $M$, and $M'$, where $M'$ is (like $M$) assumed to be a classical system, in this case a bit, with observable $Z_{M'}$. Two, that the conservation law holds for the quantity $Z_Q+Z_M+Z_{M'}$. This corresponds to a situation where the channel on $Q$ and $M$ is generated by interacting with an effectively classical environment. If on the other hand $Q$ is allowed to interact coherently with another quantum system in the environment, the latter could be mediating the coherent evolution instead of $M$, so the temporal witness would no longer be applicable. This would be analogous to a situation where, in the gravitational witness, the two quantum probes were allowed to interact via another quantum system that is not the mediator. In that case the witness does not allow one to conclude anything about the mediator; the same is true in the case of the temporal witness.

Note that the conservation law is crucial to reach our conclusion. Without it, one could allow for more general unitaries involving observables of $Q$ that do not commute with $Z_Q$, hence effectively allowing for a non-trivial quantum-coherent dynamics on the qubit $Q$, occurring even if $M$ is classical. For instance one could set $Q$ to the state $\ket{+}$ or $\ket{-}$, starting from the state $\ket{0}$, according to whether the classical variable $Z_M$ is sharp with value $0$ or $1$.

Now assuming that $M$ is a qubit, we can formally write the allowed Hamiltonians that comply with the conservation law and can perform the witnessing task. By imposing condition \eqref{CON}, we can see that the only allowed unitaries are generated by the Hamiltonian:
\begin{align}
H_{MQ}=& a Z_Q +b Z_M +c Z_QZ_M +f \left (X_QX_M+Y_QY_M \right )& \nonumber \\ 
& +g \left (X_QY_M-Y_QX_M\right ) +t {\mathcal{I}} \nonumber
\end{align}
where $\mathcal{I}$ is the unit defined on the full two-qubit space; $A_{\alpha}$ represents the component $A\in\{X, Y, Z\}$ of the qubit $\alpha\in \{Q, M\}$.

An elementary unitary that complies with this requirement is the \SWAP\ gate. Assuming that $M$ is already in an eigenstate of $X_M$, we can see how by applying the \SWAP\ gate can prepare $Q$ in an eigenstate of $X_Q$. This of course requires $M$ to be non-classical as it must be preparable in an eigenstate of the variable $X_M$,  which is incompatible with $Z_M$.
More general Hamiltonians allowed by the conservation law constraint include also the exchange interaction:$$H\propto (S_{Q+}S_{M-}+S_{Q-}S_{M+})$$ where $S_{Q\pm}=(X_Q\pm i Y_Q)$ (and similarly for $M$).
In all these $M$ engages non-commuting degrees of freedom to mediate the non-trivial coherent evolution on $Q$, as expected given our result. 

\subsection{Conservation of a non-additive quantity}

We shall now consider a more general conservation law, which mandates the conservation of the non-additive quantity $Z_Q+Z_M+Z_QZ_M$. Under this constraint, the only allowed unitaries $U_{MQ}$ must satisfy:
\begin{equation}
    [Z_Q+Z_M+Z_QZ_M, U_{MQ}]=0
    \label{conINT}
\end{equation}
It is straightforward to show that assuming this conservation law leads to the same argument as when assuming equation \eqref{CON}. The generalisation to a quantum channel operating on $Q$ and $M$ holds too, under the appropriate assumptions following \eqref{conINT}. A formal proof is given in the appendix. We now turn to discussing a toy model performing the witnessing task and satisfying this condition \eqref{conINT}.

\subsubsection{Toy model with two qubits} 
We shall now discuss a sequence of unitary gates that satisfy equation \eqref{conINT} and can induce the witnessing task on $Q$, by interaction with $M$ only. This is a \textit{toy model} illustrating the temporal witness -- note however that our general argument in support of the witness did not assume any specific dynamics. To do so, we shall adopt the Heisenberg picture. 

\begin{adjustbox}{width=\columnwidth}
\begin{quantikz}[slice all, remove end slices=1, slice titles=$t_\col$]
\lstick{Q} & \targ{} & \qw & \control{} & \swap{1} & \qw & \targ{} & \qw  \\
\lstick{M} & \ctrl{-1} & \gate{R_Y^{\left(\frac{\pi}{2}\right)}} & \ctrl{-1} & \targX{} & \gate{R_Y^{\left(-\frac{\pi}{2}\right)}} & \ctrl{-1} & \qw 
\end{quantikz}
\end{adjustbox}

Let $\hat{q}^{(Q)}\coloneqq\left(\sigma_x \otimes \mathcal{I}_{M}, \sigma_y \otimes \mathcal{I}_{M}, \sigma_z \otimes \mathcal{I}_{M}\right)$, where $\sigma_{\alpha}$, $\alpha=x,y,z$ are the Pauli operators, be the vector of generators $q_{\alpha}^{(Q)}$ of the algebra of observables of the qubits $Q$. Let $\hat{q}^{(M)}$ be defined in a similar way. A quantum network that can perform the witnessing task is represented below. 
First one applies a controlled-not (\CNOT) gate using $M$ as control qubit; one then applies a rotation of an angle $\theta=\frac{\pi}{2}$ about the Y-axis; then a controlled-phase (\CPHASE) gate, so that the qubits become entangled independently of the initial state of $M$; a \SWAP\ gate follows; finally, the same sequence of rotation ($\theta'=-\frac{\pi}{2}$) and \CNOT\ gates is applied to the system in order to disentangle the qubits. The gates have the following expressions:
\begin{align}
    \CNOT_{M,Q}(t_i)=&\frac{1}{2}\left(\mathcal{I}+q_z^{(M)}(t_{i-1})\right) \nonumber & \\
    &+\frac{1}{2}\left(\mathcal{I}-q_z^{(M)}(t_{i-1})\right)q_x^{(Q)}(t_{i-1})&
    \label{CNOThamiltonian}
\end{align}
\begin{align}
    \CPHASE_{M,Q}(t_i)=&\frac{1}{2}\left(\mathcal{I}+q_z^{(M)}(t_{i-1})\right)\nonumber& \\ &+\frac{1}{2}\left(\mathcal{I}-q_z^{(M)}(t_{i-1})\right)q_z^{(Q)}(t_{i-1})&
    \label{CPHhamiltonian}
\end{align}
\begin{equation}
    \R_{Y_{(\pm \frac{\pi}{2})}}^{(M)}(t_i)=\frac{\sqrt{2}}{2}\left(\mathcal{I} \mp i q_y^{(M)}(t_{i-1})\right)
    \label{ROThamiltonian}
\end{equation}
\begin{align}
    \SWAP(t_i)=&\frac{1}{2}\left(\mathcal{I}+q_x^{(Q)}(t_{i-1})q_x^{(M)}(t_{i-1})+ \nonumber \right. & \\ & \left. +q_y^{(Q)}(t_{i-1})q_y^{(M)}(t_{i-1})+q_z^{(Q)}(t_{i-1})q_z^{(M)}(t_{i-1})\right) &
    \label{SWAPhamiltonian}
\end{align} and the final unitary for the network is:
\begin{align}
    U_{net}=&\CNOT_{M,Q}\cdot\R^{(M)}_{Y_{(-\frac{\pi}{2})}}\cdot\SWAP\cdot\CPHASE_{M,Q}\cdot & \nonumber \\ & \cdot\R^{(M)}_{Y_{(\frac{\pi}{2})}}\cdot\CNOT_{M,Q}.
    \label{definitiveHamiltonian}
\end{align}
The evolution of the Heisenberg descriptors $\left\{q_x^{(Q,M)}, q_y^{(Q,M)}, q_z^{(Q,M)}\right\}$ is shown in the Table \ref{tab:evolutiondescriptorscircuit}.\\
\begin{table}
    \centering
    \resizebox{\columnwidth}{!}{%
    \begin{tabular}{c|c|c}
        & Qubit Q & Qubit M \\
        \hline
         $t_0$ & $\left\{q_x^{(Q)},q_y^{(Q)},q_z^{(Q)}\right\}$ & $\left\{q_x^{(M)},q_y^{(M)},q_z^{(M)}\right\}$ \\
         $t_1$ & $\left\{q_x^{(Q)},q_y^{(Q)}q_z^{(M)},q_z^{(Q)}q_z^{(M)}\right\}$ & $\left\{q_x^{(Q)}q_x^{(M)},q_x^{(Q)}q_y^{(M)},q_z^{(M)}\right\}$ \\
         $t_2$ & $\left\{q_x^{(Q)},q_y^{(Q)}q_z^{(M)},q_z^{(Q)}q_z^{(M)}\right\}$ & $\left\{q_z^{(M)},q_x^{(Q)}q_y^{(M)},-q_x^{(Q)}q_x^{(M)}\right\}$ \\
         $t_3$ & $\left\{-q_x^{(M)},-q_z^{(Q)}q_y^{(M)},q_z^{(Q)}q_z^{(M)}\right\}$ & $\left\{q_z^{(Q)},q_y^{(Q)}q_x^{(M)},-q_x^{(Q)}q_x^{(M)}\right\}$ \\
         $t_4$ & $\left\{q_z^{(Q)},q_y^{(Q)}q_x^{(M)},-q_x^{(Q)}q_x^{(M)}\right\}$ & $\left\{-q_x^{(M)},-q_z^{(Q)}q_y^{(M)},q_z^{(Q)}q_z^{(M)}\right\}$ \\
         $t_5$ & $\left\{q_z^{(Q)},q_y^{(Q)}q_x^{(M)},-q_x^{(Q)}q_x^{(M)}\right\}$ & $\left\{-q_z^{(Q)}q_z^{(M)},-q_z^{(Q)}q_y^{(M)},-q_x^{(M)}\right\}$ \\
         $t_6$ & $\left\{q_z^{(Q)},-q_y^{(Q)},q_x^{(Q)}\right\}$ & $\left\{-q_z^{(M)},-q_y^{(M)},-q_x^{(M)}\right\}$ \\ 
    \end{tabular}
    }
    \caption{Evolution of the system $Q\oplus M$ in the Heisenberg Picture. All the components are expressed as function of the descriptors at time $t_0$.}
    \label{tab:evolutiondescriptorscircuit}
\end{table}
Notice in particular that the proposed unitary satisfies the condition \eqref{conINT} and the system $Q$ cannot evolve without interacting with $M$. Moreover the Heisenberg picture allows us to see directly that the witnessing task is successfully performed independently of the initial state of the mediator $M$.

\subsubsection{Toy model with a system of \textit{N} qubits}

We shall now discuss a more general case where $M$ is made of $N$ qubits as mediators of the witnessing task on the system qubit $Q$. This can be done exploiting the \textit{Quantum Homogeniser} \cite{HOMOGENISER},  a  quantum  device  able  to  “homogenise”  the quantum state of the system with the state of the reservoir of $N$ qubits to an arbitrary precision improving as we increase $N$. 

The initial state for the system is $\rho_0=\frac{1}{2}(\mathcal{I}+q_z^{(Q)})$, while we prepare the homogeniser made of $N$ qubits in the state $\xi=\frac{1}{2}(\mathcal{I}+q_x^{(M)})$ , so that:
\begin{equation}
    (\rho\otimes\xi^{\otimes N})_0=\frac{1}{2}(\mathcal{I}+q_z^{(Q)})\otimes\frac{1}{2}(\mathcal{I}+q_x^{(M)})^{\otimes N}.
\end{equation}
The system qubit interacts with each qubit in the reservoir (one by one and only once) at each time step. The interaction is described by the unitary \textit{partial \SWAP} gate:
\begin{equation}
    P(\eta)=\cos{\eta}\mathcal{I}+i\sin{\eta} S
\end{equation} where $S$ is the \SWAP\ gate in \eqref{SWAPhamiltonian} and $\eta$ represents the strength of the homogenisation. It can be shown that this is the \textit{unique} unitary operation able to perform the homogenisation we are looking for \cite{HOMOGENISER}. Notice moreover that $P(\eta)$ satisfies the condition \eqref{conINT}.\\
After $n$ interactions, we obtain:
\begin{eqnarray}
    &\rho^{(n)}=\cos^2\eta{\rho^{(n-1)}}+\sin^2\eta{\xi}+i\cos{\eta}\sin{\eta}\left[\xi,\rho^{(n-1)}\right] &  \\ &
    \xi'_n=\cos^2\eta{\xi}+\sin^2\eta{\rho^{(n-1)}}-i\cos{\eta}\sin{\eta}\left[\xi,\rho^{(n-1)}\right] &
\end{eqnarray} which can be rewritten focusing on the terms proportional to $\xi$:
\begin{equation}
    \rho^{(n)}=\sin^2{\eta}\sum_{k=0}^{n-1}\cos^{2k}\eta{\xi}+\rho^{(n)}_{rest}=\left(1-\cos^{2n}{\eta}\right)\xi + \rho^{(n)}_{rest}
\end{equation}
\begin{equation}
    \xi'_n=\sin^2{\eta}\left(1-\cos^{2(n-1)}{\eta}\right)\xi+\xi_{n,rest}.
\end{equation}
where $\xi'_n$ is the state of the reservoir qubits after $n$ interactions with the system one and $\rho^{(n)}_{rest}$ and $\xi_{n,rest}$ the components of the density matrices of the system and reservoir qubit respectively that do not depend on the initial state of the reservoir qubits $\xi$.
Since the homogenisation is a \textit{contractive} map \cite{HOMOGENISER}, $\rho^{n}_{rest}$ converges monotonically to $\emptyset$, meaning that $\rho^{(N)}\rightarrow\xi$ with an accuracy increasing with the number of qubits $N$ in the reservoir. The non-classicality of the mediator is, in this case, associated with the interaction between $M$ and $Q$ (which involves all components of $M$, which are incompatible variables) as well as with the state that $M$ is initialised in. 

\subsubsection{Toy model with a harmonic oscillator}
Finally, we provide a toy model where $M$ is a harmonic oscillator (with creation and annihilation operators denoted by $b^\dagger$ and $b$).
We start from the unitary in \eqref{definitiveHamiltonian}, introduced for the two qubits scenario, and apply the \textit{Holstein-Primakoff transformation} \cite{H-P}:
\begin{equation}
    \begin{cases}
        S_z=\hbar\left(s-a^\dagger a\right) \\
        S^+=\hbar\sqrt{2s}\sqrt{1-\frac{a^\dagger a}{2s}}a \\
        S^-=\hbar\sqrt{2s}a^\dagger\sqrt{1-\frac{a^\dagger a}{2s}}
    \end{cases}
\end{equation} being $s$ the particle's spin, $\hat{S}=\left(S_x,S_y,S_z\right)$ the spin operator and $S^{\pm}=S_x\pm i S_y$ the raising and lowering operators. We can describe the qubit as a spin $s=\frac{1}{2}$ particle so that, reintroducing the usual descriptors $\hat{q}=\left(q_x,q_y,q_z\right)$ as above, we get ($\hbar=1$ from now on):
\begin{equation}
    \begin{cases}
        q_x=\frac{\sqrt{1-a^\dagger a}a + a^\dagger\sqrt{1-a^\dagger a}}{2} \\
        q_y=\frac{\sqrt{1-a^\dagger a}a - a^\dagger\sqrt{1-a^\dagger a}}{2i} \\
        q_z=\frac{1}{2}-a^\dagger a.
    \end{cases}
    \label{eq:H-P transformation}
\end{equation}

%The detailed description of this transformations are discussed in the Appendix; 
Applying this transformation to \eqref{definitiveHamiltonian}, we find that the unitary for the toy model is:
\begin{align}
    U=\frac{1}{2}&\left[\left(-\frac{\sqrt{1-b^\dagger b}b + b^\dagger\sqrt{1-b^\dagger b}}{2}+\frac{1}{2}-b^\dagger b\right)\cdot \right. & \nonumber  \\ & \left. \cdot \left( \frac{\sqrt{1-a^\dagger a}a + a^\dagger\sqrt{1-a^\dagger a}}{2}+\frac{1}{2}-a^\dagger a\right)\right]
    \label{eq:definitiveHamiltonianH-P}
\end{align} where we have neglected the constant terms.
In \eqref{eq:definitiveHamiltonianH-P} the terms $\hbar\omega_a a^\dagger a$ and $\hbar\omega_b b^\dagger b$ describe the free evolution of our subsystems; while the three interaction terms suggest that the harmonic oscillator is a controller for the spin-$\frac{1}{2}$ particle.
Moreover, the non-commuting, incompatible degrees of freedom of the mediator are the number operator $b^\dagger b$ and the creation and annihilation operators $b^\dagger$ and $b$. They are involved in the interaction that makes the witnessing task possible on $Q$, which concludes our final generalisation to the continuous variable limit for the control system $M$. Whichever the dimensionality of the control system's Hilbert space, if it is able to induce a coherent rotation on the qubit, then it \textit{must} be non-classical, provided a global quantity on the system $Q\oplus M$ is conserved.

We finally note that a straightforward way to generalise the entanglement-based witness to the temporal case is to consider that if we represent $Q$ in second-quantisation, the coherent rotation of $Q$ from one basis to some other non-commuting basis corresponds to the creation of some mode-entanglement on $Q$ \cite{mode-ent}. If this entanglement is mediated via $M$ (and it must be so, because of the conservation law) then using the entanglement-based witness we can conclude directly that $M$ is non-classical. Again, we see that the conservation law in this case is a sufficient condition that implies the impossibility of building a gate that entangles the two modes in $Q$ directly. So in this particular case it can be used as a substitute to the locality requirement featuring in \cite{MAVE}. 

\section{Discussion}
We have proposed a generalisation of the entanglement-based witness of non-classicality of a system $M$, extending it to the temporal case, with a single quantum probe interacting with $M$. 
One of the interesting implications of this work is that to achieve a temporal version of the witness we need to add a conservation law to the set of assumptions. Standard conservation laws (such as those for charge, energy and momentum) are usually assumed as background knowledge for experimental tests, as they have been verified independently. In our toy model, we considered both additive and the non-additive conservation laws, the former coherently with central tenets of fundamental laws (e.g. energy and momentum conservation laws), the latter to generalise our argument to putative, more general, conservation laws. In a real case, following quantum field theory, the stress-energy tensor conservation law seems a plausible candidate. This constraint plays the same role as the requirement of local interactions in the entanglement-based version of the witness, where the two quantum probes cannot interact directly but only, locally, via the third mediating system. This is the reason why $Q$ cannot evolve on its own, but it requires the assistance of $M$ which is a mediator of $Q$'s evolution in time. 

This leads to an interesting conjecture about the meaning of ``locality in time'': the conservation law plays the same role of the requirement of local interactions between the probes and the mediator $M$, i.e., the requirement of \textit{locality in space}; so, by analogy, it can be considered as requiring \textit{locality in time}. Locality in time, in our case, corresponds to the fact that the evolution of a system from one state to a different state cannot occur unless when mediated by the interaction with another system, the mediator.
%\textcolor{blue}{the time evolution of a physical system is said to be \textit{local} when it is influenced by the state of a different system at the \textit{same} time, meaning that it cannot evolve on its own, without interacting with the latter}. 
Hence it may be possible to find a conservation law in the spatial form of the witness, which can substitute the requirement of locality in space in the same way as the conservation law introduced in the temporal argument achieves the requirement of locality in time. This conjecture will be explored in future works.

We note here that the test with a single mass described in \cite{HOW} is of a different nature: it aims at distinguishing two different models for gravity, one classical and one quantum. If the quantum model is borne out in an experiment, we can reject the particular classical model for gravity. However, other classical models for gravity are not excluded. Our argument aims to be more general, as it excludes {\sl all} classical models for the system $M$ which induces the interaction and comply with our assumptions. In order to achieve full generality and emancipate this argument fully from the formalism of quantum theory, one has to express the assumptions in strict information-theoretic terms,  like in \cite{MRVPRD}, without relying on unitary quantum theory's formalism. We shall leave this to a forthcoming paper.

{\bf Acknowledgements} \;\;We thank Vlatko Vedral for comments on earlier drafts of this manuscript. This research was made possible through the generous support of the Gordon and Betty Moore Foundation, the Eutopia Foundation and of the ID 62312 grant from the John Templeton Foundation, as part of the \href{https://www.templeton.org/grant/the-quantuminformation-structure-ofspacetime-qiss-second-phase}{‘The Quantum Information Structure of Spacetime’ Project (QISS)}. The opinions expressed in this project/publication are those of the author(s) and do not necessarily reflect the views of the John Templeton Foundation. G.D. thanks Politecnico di Torino, the Clarendon Fund and the Oxford-Thatcher Graduate Scholarship for supporting this research.

\maketitle
%\bibliography{biblio}% Produces the bibliography via BibTeX.

\begin{thebibliography}{0}%
\makeatletter
\providecommand \@ifxundefined [1]{%
 \@ifx{#1\undefined}
}%
\providecommand \@ifnum [1]{%
 \ifnum #1\expandafter \@firstoftwo
 \else \expandafter \@secondoftwo
 \fi
}%
\providecommand \@ifx [1]{%
 \ifx #1\expandafter \@firstoftwo
 \else \expandafter \@secondoftwo
 \fi
}%
\providecommand \natexlab [1]{#1}%
\providecommand \enquote  [1]{``#1''}%
\providecommand \bibnamefont  [1]{#1}%
\providecommand \bibfnamefont [1]{#1}%
\providecommand \citenamefont [1]{#1}%
\providecommand \href@noop [0]{\@secondoftwo}%
\providecommand \href [0]{\begingroup \@sanitize@url \@href}%
\providecommand \@href[1]{\@@startlink{#1}\@@href}%
\providecommand \@@href[1]{\endgroup#1\@@endlink}%
\providecommand \@sanitize@url [0]{\catcode `\\12\catcode `\$12\catcode
  `\&12\catcode `\#12\catcode `\^12\catcode `\_12\catcode `\%12\relax}%
\providecommand \@@startlink[1]{}%
\providecommand \@@endlink[0]{}%
\providecommand \url  [0]{\begingroup\@sanitize@url \@url }%
\providecommand \@url [1]{\endgroup\@href {#1}{\urlprefix }}%
\providecommand \urlprefix  [0]{URL }%
\providecommand \Eprint [0]{\href }%
\providecommand \doibase [0]{http://dx.doi.org/}%
\providecommand \selectlanguage [0]{\@gobble}%
\providecommand \bibinfo  [0]{\@secondoftwo}%
\providecommand \bibfield  [0]{\@secondoftwo}%
\providecommand \translation [1]{[#1]}%
\providecommand \BibitemOpen [0]{}%
\providecommand \bibitemStop [0]{}%
\providecommand \bibitemNoStop [0]{.\EOS\space}%
\providecommand \EOS [0]{\spacefactor3000\relax}%
\providecommand \BibitemShut  [1]{\csname bibitem#1\endcsname}%
\let\auto@bib@innerbib\@empty
%</preamble>
\end{thebibliography}%


\begin{thebibliography}{31}%
\bibitem{DEWITT} DeWitt, B. S., and Bryce Seligman, D. (2003). The global approach to quantum field theory (Vol. 114). Oxford University Press, USA.
\bibitem{WALLACE} Wallace, D. (2012). The emergent multiverse: Quantum theory according to the Everett interpretation. Oxford University Press.
\bibitem{SCH} Schrödinger, E. (1935). The present status of quantum mechanics. Die Naturwissenschaften, 23(48), 1-26.
\bibitem{WIG} Wigner, E. P. (1995). Remarks on the mind-body question. In Philosophical reflections and syntheses (pp. 247-260). Springer, Berlin, Heidelberg.
\bibitem{DEU} Deutsch, D. (1985). Quantum theory as a universal physical theory. International Journal of Theoretical Physics, 24(1), 1-41.
\bibitem{BOHR} Bohr, N., Rosenfeld, L., Kgl. Danske Vidensk. Selsk., Mat. Fys. Med. 12, 8, 1933.
\bibitem{PENROSE} Penrose, R. (1996). On gravity's role in quantum state reduction. General relativity and gravitation, 28(5), 581-600.
\bibitem{GRW} Ghirardi, G. C., Rimini, A., \& Weber, T. (1986). Unified dynamics for microscopic and macroscopic systems. Physical review D, 34(2), 470.
\bibitem{MRVPRD}  Marletto, C., and Vedral, V., Phys. Rev. D {\bf 102}, 086012,  2020.
\bibitem{SOUG} Bose, S., {\it et al.} Phys. Rev. Lett. {\bf 119}, 240401, 2017.
\bibitem{MAVE}  Marletto, C., and Vedral, V., Phys. Rev. Lett. {\bf 119}, 240402, 2017.
\bibitem{Vedral} Zhang, T., Dahlsten, O., \& Vedral, V. (2020). Quantum correlations in time. arXiv preprint arXiv:2002.10448.
\bibitem{Brukner} Brukner, C., Taylor, S., Cheung, S., \& Vedral, V. (2004). Quantum entanglement in time. arXiv preprint quant-ph/0402127.
\bibitem{REF1} Nowakowski, M., Quantum entanglement in time, AIP Conf. Proc. 1841, 020007 (2017).
\bibitem{REF2} Cotler, J., Duan, L.-M., Hou, P.-Y., Wilczek, F., Xu, D., Yin, Z.-Q., and Zu, C., Experimental test of entangled histories, Ann. Phys. 387, 334 (2017).
\bibitem{REF3} Nowakowski, M., Cohen, E., and Horodecki, P.,  Entangled histories versus the two-state-vector formalism: Towards a better understanding of quantum temporal correlations, Phys. Rev. A 98, 032312 (2018).
\bibitem{Leggett} Leggett, A. J., and Garg, A. (1985). Quantum mechanics versus macroscopic realism: Is the flux there when nobody looks?. Physical Review Letters, 54(9), 857.
\bibitem{HOW} Howl, R., et al. "Non-Gaussianity as a signature of a quantum theory of gravity." PRX Quantum 2.1 (2021): 010325.
%\bibitem{SOUG2} R. Marshman, {\sl et al.} Phys. Rev. A 101, 052110, 2020.
%\bibitem{LOCC} R. Horodecki et al., Rev. Mod. Phys.81:865-942, 2009.
%\bibitem{HARE} M. Hall and M. Reginatto, J. Phys. A: Math. Theor. 51, 085303, 2018.
%\bibitem{Collapse} G. Ghirardi and A. Bassi, The Stanford Encyclopedia of Philosophy (Summer 2020 Edition), Edward N. Zalta (ed.), 
%\bibitem{DYS} T. Rothman and S. Boughn, Found Phys 36, 1801-1825 (2006). 
%\bibitem{PAL2021} S. Pal {\sl et al.}, Quantum 5, 478, 2021.  
%\bibitem{TER1} A. Ahmadzadegan, R. Mann, D. R. Terno, Phys. Rev A, {\bf 93}, 2016.
\bibitem{MRVNPJ} Marletto, C., and Vedral, V., npj Quantum Information \textbf{3}, 29 (2017).
\bibitem{MRVARXIV} Marletto, C., Vedral, V. Witnessing the quantumness of a system by observing only its classical features. npj Quantum Inf 3, 41 (2017).
\bibitem{CLAS} Kafri, D., Taylor, J. M., and Milburn, G. J., New J. Phys. 16, 065020 (2014).
\bibitem{DEMA} Deutsch, D., and Marletto, C., Proc. R. Soc. A {\bf 471}, 2174, 2014.
%\bibitem{BARCE} C. Barcelo {\sl et al.}, Phys. Rev. A 86, 042120, 2012.
%\bibitem{TER} A. Peres and D. Terno, Phys. Rev. A 63, 2001. 
%\bibitem{ELZE} H.T. Elze, Phys. Rev. A 85, 052109, 2012.
%\bibitem{ELZE2} L. Fratino {\sl et al.}, Phys. Scr. T 163 014005, 2014.
%\bibitem{HARE-Book} M. Hall and M. Reginatto, Ensembles on Configuration Space, Springer International Publishing (2016), Chapters 3-9.
%\bibitem{SAL} L. L. Salcedo, Phys.Rev. A85, 022127, 2012.
%\bibitem{OLY} T. A. Olyinyk, Foundations of Physics volume 46, 2016.
%\bibitem{SUD3} G. Sudarshan {\sl et al.}, Phys.Rev.D 20, 3081-3093, 1970.
%\bibitem{Harvey-Bohm} M. Redhead and H. Brown,
%Proceedings of the Aristotelian Society, Supplementary Volumes, Vol. 65 (1991), pp. 119-159.
\bibitem{HOMOGENISER} Ziman, M., et al. "Quantum homogenization." arXiv preprint quant-ph/0110164 (2001).
\bibitem{H-P} Holstein, T., and Hl Primakoff. "Field dependence of the intrinsic domain magnetization of a ferromagnet." Physical Review 58.12 (1940): 1098.
\bibitem{mode-ent} Cunha, M. O. T., Dunningham, J. A., \& Vedral, V. (2007). Entanglement in single-particle systems. Proceedings of the Royal Society A: Mathematical, Physical and Engineering Sciences, 463(2085), 2277-2286.


\end{thebibliography}


\section{APPENDIX} 
\subsection{The general temporal argument}
We want to show that the condition for the control system to be quantum is also \textit{necessary}, by proving the following theorem:
\begin{teorema}
If a physical system $Q$ evolves from an eigenstate of one of its observables to an eigenstate of a different, non-commuting observable under the conservation law of a global quantity of the system $Q\oplus M$, then $M$ must be non-classical.
\end{teorema}
The proof will go by contradiction. Let us suppose, for simplicity, that the system $Q$ is a qubit described by the descriptors $\hat{q}^{(Q)}=\left(\sigma_x \otimes \mathcal{I}_M, \sigma_y \otimes \mathcal{I}_M, \sigma_z \otimes \mathcal{I}_M\right)=\left(q_x^{(Q)}, q_y^{(Q)}, q_z^{(Q)}\right)$ while the system $M$ is a \textit{classical} bit described by the z-component of its descriptors vector $q_z^{(M)}=\mathcal{I}_Q \otimes \sigma_z$.\\
The system $Q$ is initialised in a generic state $\ket{\psi}_{0}$ and we suppose that the physical quantity $q_z^{(Q)}+q_z^{(M)}+q_z^{(Q)}q_z^{(M)}$ is conserved, so that $Q$ cannot evolve if isolated, it can evolve only through its interaction with $M$. 
The initial density matrix will read:
\begin{eqnarray}
    &\rho_0=\frac{1}{4}\left(\mathcal{I}_{Q,M}+r_x q_x^{(Q)}+ r_y q_y^{(Q)} + r_z q_z^{(Q)} + s_z q_z^{(M)} \right. \nonumber &
    \\ & \left. + t_x q_x^{(Q)}q_z^{(M)} + t_y q_y^{(Q)}q_z^{(M)} + t_z q_z^{(Q)}q_z^{(M)}\right)&
\end{eqnarray}being $\Vec{r}=\left(r_x, r_y, r_z\right)$, $\Vec{t}=\left(t_x,t_y,t_z\right)$ and $s_z$ real valued coefficients. We let it evolve under the most general Hamiltonian describing the interaction between $Q$ and $M$:
\begin{eqnarray}
    &H_{QM}=\alpha q_x^{(Q)} + \beta q_y^{(Q)} + \gamma q_y^{(Q)} + \nonumber & \\
    & a q_x^{(Q)}q_z^{(M)} + b q_y^{(Q)}q_z^{(M)} + c q_z^{(Q)}q_z^{(M)}. &
    \label{eq:mostgeneralHamiltonian}
\end{eqnarray}Before doing this, we have to enforce the conservation law required for $Q$ to be unable to evolve spontaneously:
\begin{equation}
    \left[H_{QM}, q_z^{(Q)}+q_z^{(M)}+q_z^{(Q)}q_z^{(M)}\right]=0
\end{equation}which implies:
\begin{eqnarray}
    &\alpha=-a \nonumber& \\
    &\beta=-b \nonumber&
\end{eqnarray}so that:
\begin{eqnarray}
    &H'_{QM}=\alpha q_x^{(Q)} + \beta q_y^{(Q)} + \gamma q_z^{(Q)} \nonumber& \\
    & - \alpha q_x^{(Q)}q_z^{(M)} - \beta q_y^{(Q)}q_z^{(M)} + c q_z^{(Q)}q_z^{(M)}&
    \label{Hamiltonianconservedconstraint}
\end{eqnarray}and we will label $H'_{QM}$ as $H_{QM}$ to lighten the notation.

We first of all notice that $\left[H_{QM}, q_z^{(M)}\right]=0$, which means that:
\begin{equation}
    q_z^{(M)}(t)=e^{-iH_{QM}}q_z^{(M)}e^{iH_{QM}}=q_z^{(M)}
\end{equation} i.e. $M$ cannot evolve because of the interaction with $Q$: if it is $\ket{0}$, it stays in $\ket{0}$. Hence, the effects of $H_{QM}$ on $Q$ is to \textit{rotate it} around the vector whose components are:
\begin{equation}
    \begin{cases}
    n_x=\alpha(1-q_z^{(M)}) \\
    n_y=\beta(1-q_z^{(M)}) \\
    n_z=\gamma+cq_z^{(M)}.
    \end{cases}
    \label{directors}
\end{equation}
Our goal now is to find $(n_x, n_y, n_z)$ such that it is possible to perform a rotation of the qubit $Q$ mapping $q_z^{(Q)} \rightarrow q_x^{(Q)}$, $q_y^{(Q)} \rightarrow -q_y^{(Q)}$, $q_x^{(Q)} \rightarrow q_z^{(Q)}$ involving the bit $M$.\\
Looking at the components in \eqref{directors}, we notice that if the eigenvalue of $q_z^{(M)}$ is $+1$, then both $n_x$ and $n_y$ will vanish, meaning that if the bit $M$ is initialised in $\ket{0}$, $Q$ can only rotate around the z-axis, which is not what the mapping discussed above should achieve. Hence, $M$ should be initialised in $\ket{1}$, so that:
\begin{equation}
    \begin{cases}
    n_x=2\alpha \\
    n_y=2\beta \\
    n_z=\gamma-c.
    \end{cases}
    \label{directorZ-1}
\end{equation} 
Considering the unitary:
\begin{equation}
    R_\theta=\cos{\left(\frac{\theta}{2}\right)}\mathcal{I}-i\sin{\left(\frac{\theta}{2}\right)}\left(n_x q_x^{(Q)}+n_y q_y^{(Q)}+ n_z q_z^{(Q)}\right)
    \label{rotationgate}
\end{equation} with $\theta=\frac{\pi}{2}$, we can evolve the descriptors for the qubit $Q$ according to \eqref{rotationgate}:
\begin{equation}
    q_z^{(Q)}(t)=R_\theta^\dagger q_z^{(Q)} R_\theta \implies \begin{cases}
    -n_y+n_x n_z=1 \\
    n_x+n_y n_z=0 \\
    \frac{1}{2}+\frac{1}{2}\left(n_z^2-n_x^2-n_y^2\right)=0\;. 
    \end{cases}
\end{equation} This gives three roots out of which only one can be accepted because of the condition $n_x^2 + n_y^2 + n_z^2 =1$: $(0,-1,0)$.
Let us focus on the x component:
\begin{equation}
    q_x^{(Q)}=R_\theta^\dagger q_x^{(Q)} R_\theta \implies \begin{cases}
    n_y + n_x n_z = 1 \\
    -n_z + n_x n_y =0 \\
    \frac{1}{2}+ \frac{1}{2}\left(n_x^2-n_y^2-n_z^2\right)=0
    \end{cases}
\end{equation} out of which we extract three roots with a single one acceptable: $(0,1,0)$, which is consistent with what we found earlier.
Finally, we evolve the y component:
\begin{equation}
    q_y^{(Q)}=R_\theta^\dagger q_y^{(Q)} R_\theta \implies \begin{cases}
    -n_x + n_y n_z =0 \\
    n_z + n_x n_y=0 \\
    \frac{1}{2}+\frac{1}{2}\left(n_y^2 - n_x^2 - n_z^2\right)= 1 
    \end{cases}
\end{equation} which, instead, gives no acceptable roots: we cannot find a consistent axis around which the rotation we are looking for can be performed. The conclusion, as promised, is that the {witnessing} task is \textit{not possible} if the system $M$ has no non-commuting degrees of freedom.

We can generalise this discussion to a quantum channel operating on $Q$ and $M$ as in the case of the additive conservation law discussed in the main text, simply changing the second assumption according to the non-additive conservation law in \eqref{conINT}: the conserved quantity here is $q_z^{(Q)} + q_z^{(M)} + q_z^{(M')} + q_z^{(Q)}q_z^{(M')} + q_z^{(M)}q_z^{(M')}$. 

The most general Hamiltonian describing the interaction between $Q$ and $M'$, and $M$ and $M$ will be:
\begin{eqnarray}
     &H_{QM', MM'}=\alpha q_x^{(Q)} + \beta q_y^{(Q)} + \gamma q_y^{(Q)} +  a q_x^{(Q)}q_z^{(M')} \nonumber & \\ & + b q_y^{(Q)}q_z^{(M')} + c q_z^{(Q)}q_z^{(M')} + a' q_z^{(M)}q_z^{(M')} &
\end{eqnarray} which under the conservation law becomes:
\begin{eqnarray}
    &H_{QM', MM'}=\alpha q_x^{(Q)} + \beta q_y^{(Q)} + \gamma q_z^{(Q)}  - \alpha q_x^{(Q)}q_z^{(M')} \nonumber& \\ & - \beta q_y^{(Q)}q_z^{(M')} + c q_z^{(Q)}q_z^{(M')} + a' q_z^{(M)}q_z^{(M')}.&
    \label{eq:HAMQUANTUMCHANNEL}
\end{eqnarray}

Comparing \eqref{eq:HAMQUANTUMCHANNEL} with \eqref{Hamiltonianconservedconstraint}, we see that the only difference is in the term $a' q_z^{(M)}q_z^{(M')}$, which is cannot modify the state of $M$ and $M'$, as we intended to show.

\subsection{The general argument: \textit{N+1} qubits}
We have seen that a reservoir made of \textit{quantum systems} is indeed able to perform the homogenisation task. One wants to prove it to be \textit{necessary} as well.

Let us suppose that the reservoir is made of \textit{bits}, so that it is initialised in the state:
\begin{equation}
    \xi=\ket{0}\bra{0}^{\otimes N}=\frac{1}{2}\left(\mathcal{I}+q_z^{(M)}\right)^{\otimes N},
\end{equation} since the computational basis is assumed to be $\{\ket{0},\ket{1}\}$. This means that the system qubit $Q$ will be initialised in:
\begin{equation}
    \rho=\ket{+}\bra{+}=\frac{1}{2}\left(\mathcal{I}+q_x^{(Q)}\right)
\end{equation} and the task we are looking for is an homogenisation of $\rho$ to $\xi$, i.e. a rotation for the qubit $Q$ from the state $\ket{+}$ to the state $\ket{0}$.\\
We assume, as in the quantum homogeniser, that the system qubit is allowed to interact with one bit per time step and only once. The most general unitary describing the interactions will be:
\begin{equation}
    U(\eta)=\cos{\left(\eta\right)}\mathcal{I}+i\sin{\left(\eta\right)}H_{QM}
\end{equation} being $H_{QM}$ the same introduced in \eqref{eq:mostgeneralHamiltonian}. As said above, the partial \SWAP\ does conserve the physical quantity $q_z^{(Q)}+q_z^{(M)}+q_z^{(Q)}q_z^{(M)}$; this means that we should enforce again the conservation of such quantity in our Hamiltonian. We get:
\begin{eqnarray}
    &H_{QM}=\alpha q_x^{(Q)} + \beta q_y^{(Q)} + \gamma q_z^{(Q)} & \nonumber \\ & - \alpha q_x^{(Q)}q_z^{(M)} - \beta q_y^{(Q)}q_z^{(M)} + c q_z^{(Q)}q_z^{(M)}.&
    \label{HamiltonianconservedconstraintHOMOGENISER}
\end{eqnarray}
Now we are ready to let the system qubit interact with the first \textit{bit} in the reservoir:
\begin{eqnarray}
     &\left(\rho \otimes \xi\right)^1=U^\dagger(\eta)\left(\rho\otimes\xi\right)U(\eta)  \nonumber &\\ &
     = \frac{1}{4}\cos^2{\left(\eta\right)}\left(\mathcal{I}+q_x^{(Q)}+q_z^{(M)}+q_x^{(Q)}q_z^{(M)}\right) \nonumber & \\ &
     + \frac{1}{4}\sin^2{\left(\eta\right)}\left(\mathcal{I}+H^\dagger_{QM}q_x^{(Q)}H_{QM} \right .\nonumber & \\ & \left. +q_z^{(M)}+H^\dagger_{QM}q_x^{(Q)}H_{QM}q_z^{(M)}\right) \nonumber & \\ &
     + i\sin{\left(\eta\right)}\cos{\left(\eta\right)}\left[q_x^{(Q)},H_{QM}\right] \nonumber & \\ & + i\sin{\left(\eta\right)}\cos{\left(\eta\right)}\left[q_x^{(Q)},H_{QM}\right]q_z^{(M)}.&
     \label{eq:evolutionproofNqubits}
\end{eqnarray}
Let us focus on the term proportional to $\sin^2{\eta}$: it must be proportional to $\xi\otimes\rho$ in order for the fixed point of the contractive map to be reached at the end of the homogenisation procedure. Of course, we cannot expect the bit to rotate in an eigenstate of the $X$ operator, so what we are actually looking for is a proportionality with $\xi\otimes\xi$. We have:
\begin{equation}
    \mathcal{I}+H^\dagger_{QM}q_x^{(Q)}H_{QM}+q_z^{(M)}+H^\dagger_{QM}q_x^{(Q)}H_{QM}q_z^{(M)}
\end{equation} that should match:
\begin{equation}
    \mathcal{I}+q_z^{(Q)}+q_z^{(M)}+q_z^{(Q)}q_z^{(M)},
\end{equation} but in order for this to be possible, $H_{QM}$ should be able to rotate the qubit $Q$ around a given axis such that $\ket{+}\rightarrow\ket{0}$ and we have shown in the appendix that it is \textit{not} possible to define consistently such a rotation axis.\\
This means that we will \textit{never} have a term like $\sin^2{\left(\eta\right)}\xi\otimes\xi$ in \eqref{eq:evolutionproofNqubits}, independently on the other terms that may appear in it, so that it is \textit{not possible} to perform the task we are looking for using a \textit{classical} reservoir.

\end{document}